\documentclass{PoS}

\newcommand{\ap}{Astr. Phys.}
\newcommand{\apj}{ApJ}

\usepackage{multirow}
\usepackage{siunitx}
\usepackage{comment}
\usepackage{textcomp}
\usepackage{multicol}

\title{The Cherenkov Telescope Array sensitivity to the transient sky}

\ShortTitle{CTA sensitivity to the transient sky}

\author{Valentina Fioretti$^{+,a}$, Deivid Ribeiro$^{+,b}$, Thomas B. Humensky$^{b}$, Andrea Bulgarelli$^{a}$, Gernot Maier$^{c}$, Abelardo Moralejo$^{d}$, Cosimo Nigro$^{c}$, for the CTA Consortium\footnote{for consortium list see PoS(ICRC2019)1177}\\
     \llap{$^a$} INAF Osservatorio di Astrofisica e Scienza dello Spazio di Bologna, via Gobetti 93/3, I-40129 Bologna, Italy\\
     \llap{$^b$} Department of Physics, Columbia University, 538 West 120th Street, New York, NY 10027, USA\\
     \llap{$^c$} Deutsches Elektronen-Synchrotron, Platanenallee 6, 15738 Zeuthen, Germany\\
     \llap{$^d$} Institut de Fisica d'Altes Energies, Barcelona Institute of Science and Technology, Campus UAB, 08193 Bellaterra, Barcelona, Spain\\
     \llap{$^+$} corresponding authors\\
     E-mail:  \email{valentina.fioretti@inaf.it, dr2792@columbia.edu}}

\abstract{The Cherenkov Telescope Array (CTA) will be able to perform unprecedented observations of the transient very high-energy sky. An on-line science alert generation (SAG) pipeline, with a required 30 second latency, will allow the discovery or follow-up of gamma ray bursts (GRBs) and flaring emission from active galactic nuclei, galactic compact objects and electromagnetic counterparts of gravitational waves or neutrino messengers. The CTA sensitivity for very short exposures does not only depend on the technological performance of the array (e.g. effective area, background discrimination efficiency). The algorithms to evaluate the significance of the detection also define the sensitivity, together with their computational efficiency in order to satisfy the SAG latency requirements. 
We explore the aperture photometry and likelihood analysis techniques, and the associated parameters (e.g. on-source to off-source exposure ratio, minimum number of required signal events), defining the CTA ability to detect a significant signal at short exposures. The resulting CTA differential flux sensitivity as a function of the observing time, obtained using the latest Monte Carlo simulations, is compared to the sensitivities of Fermi--LAT and current-generation IACTs obtained in the overlapping energy ranges.
}

\FullConference{36th International Cosmic Ray Conference -ICRC2019-\\
		July 24th - August 1st, 2019\\
		Madison, WI, U.S.A.}

\begin{document}

\section{Very High Energy transients in the CTA era}
The Very High Energy (VHE) domain has a great potential in unveiling both galactic (e.g. binaries) and extragalactic (e.g GRBs and active galactic nuclei) variable sources. However, current Imaging Air Cherenkov Telescopes (IACTs) are still strongly limited in sensitivity and detection on a short time scale ($\sim100$ s) are only possible for few bright sources.
The Cherenkov Telescope Array (CTA) is the future ground-based large observatory \cite{CTA1} for Very High Energies (VHE). 
The combination of Large-Sized Telescopes, operating from $\sim20$ to 150 GeV, Medium-Sized Telescopes, observing the core - 150 GeV to 5 TeV - energy range, and Small-Sized Telescopes reaching the highest energies, up to 300 TeV, will ensure a sensitivity (at 50 hr exposure) of more than one order of magnitude better than current Cherenkov experiments over a broad energy range \cite{gernot_icrc}.
CTA is currently entering the construction phase and will be operated, starting from 2022, for 30 years.
Its unprecedented scientific performance places CTA as the VHE observatory of reference for detection and monitoring of gamma-ray transients at the peak of the time-domain astronomy era. For this purpose, the observatory is equipped with a Science Alert Generation system that, by means of a real time analysis pipeline running on-site \cite{rta2013}, will be able to send alerts within 30 s from the last acquired event, with a maximum telescope positioning time of $\rm 90~s$ in response to external or internal triggers.

\section{Sensitivity at short exposures}

A fast reaction system must be coupled to a good sensitivity at very short (from seconds to hours) exposure to enable simultaneous observations of the transient sky. CTA sensitivity depends on the instrument response (e.g. effective area and angular resolution), observation conditions (e.g. night sky and cosmic-ray induced background) and the algorithms used to extract the signal and evaluate its significance. 
With CTA we will be able to collect enough signal to produce for the first time significant observations at very short ($\sim10$ s) time windows.
We base our evaluation on the standard aperture photometry method (Sec. \ref{sec:onoff}) while testing its limits and comparing it to modern likelihood analysis techniques (Sec. \ref{sec:ctools}). CTA sensitivity, at four selected energy bins, will then be computed as a function of the observation time and compared to \emph{Fermi}-LAT, MAGIC and VERITAS performance (see Sec. \ref{sec:sens_time}).
Following CTA standard rules for performance evaluation, the energy range is divided in five-per-decade equal logarithmic energy bins. The differential sensitivity is defined as the minimum flux to obtain a probability equivalent to 5 standard deviations from a point-like source in each energy bin. Unless specified otherwise, CTA results are obtained from the CTA Prod3b (Dec. 2018) Instrument Response Function (IRF) production \cite{gernot_icrc}. 

\subsection{\textit{On-Off} method}\label{sec:onoff}
A standard method in IACTs to measure the source emission is by aperture photometry, the so-called \textit{On-Off} method. The background is derived from the data by performing an observation of an \textit{Off} region with no sources in it, extracting the number of background counts N$_{\rm off}$. The background signal is then subtracted from the observation of an \textit{On} region that contains the target, with a number of counts N$_{\rm on}$. 
The two observations are usually characterized by different effective area (A) and/or exposure (t) and/or size of the region (k). The parameter $\alpha$ is given by the ratio of such parameters between the \textit{On} and the \textit{Off} regions:
\begin{equation}
\alpha = \frac{\rm A_{on}\cdot t_{on}\cdot k_{on}}{\rm A_{off}\cdot t_{off}\cdot k_{off}} \, .
\end{equation}
One may estimate the number of the background counts in the \textit{On} observation and evaluate the signal counts (N$_{\rm s}$) as follows: $\rm N_{s} = N_{on} - \alpha N_{off}$.
The significance of the detection is then computed in the IACT community using equation (17) of \cite{1983ApJ...272..317L}. This formula, based on the method of maximum likelihood ratio test, requires the N$_{\rm on}$ and N$_{\rm off}$ to be \textit{not too few} - at least 10 as rule of thumb. 

The requirements that must be satisfied
in order to declare a detection significant are: (i) a minimum number of signal counts N$_{\rm s}$ = 10 must be collected for each detection, (ii) the signal above the background must be at least five times the systematic uncertainty in the background estimation, assumed to be 1\% \cite{2013APh....43..348F}. The $\alpha$ ratio between the \textit{On} and \textit{Off} regions used in the present work is 0.2 and based on CTA standard sensitivity evaluation procedures\footnote{\href{https://www.cta-observatory.org/science/cta-performance/}{https://www.cta-observatory.org/science/cta-performance/}}.

\subsection{Full sky map maximum likelihood}\label{sec:ctools}
We used \texttt{ctools}\footnote{\href{http://cta.irap.omp.eu/ctools/}{http://cta.irap.omp.eu/ctools/}} sensitivity script \texttt{cssens} to find the best estimate of the sensitivity for the likelihood modeling technique. 
\texttt{cssens} uses a linear fit to find the approximate normalization that converges to the threshold test statistic (TS) of 25 by simulating a full sky map using the \texttt{ctobssim} method, and then fitting with \texttt{ctlike}. As for the Li\&Ma equation, the test statistic follows a $\chi^{2}$ distribution for $n$ degrees of freedom (d.o.f.): TS = $-2\, \rm ln (\lambda)$, where $\lambda = \rm L_{0}(E) / L(E)$ is the ratio between the null (only background) and alternative (background plus source) hypothesis and E is energy. The model for the source is a Crab-like power law as defined in the CTA IRFs.
The fitting procedure only varies the source's normalization. For $n=1$ d.o.f., the TS is proportional to the standard normal variable ($\sigma\sim \sqrt{\rm TS} = 5$).
The linear relation between TS and sensitivity normalization is used to estimate the input parameter for each subsequent simulation. 

In contrast to the \textit{On-Off} method, the \texttt{cssens} script (i) does not require a minimum number of source counts,
(ii) in the present implementation does not include background systematic uncertainties, (iii) takes into account the full field of view ($\sim5^{\circ}$ radius region) for the estimation of the background in the signal region and (iv) applies no angular cuts in the excess identification.
   \begin{figure}[h!]
   \centering
   \includegraphics[width=0.81\textwidth]{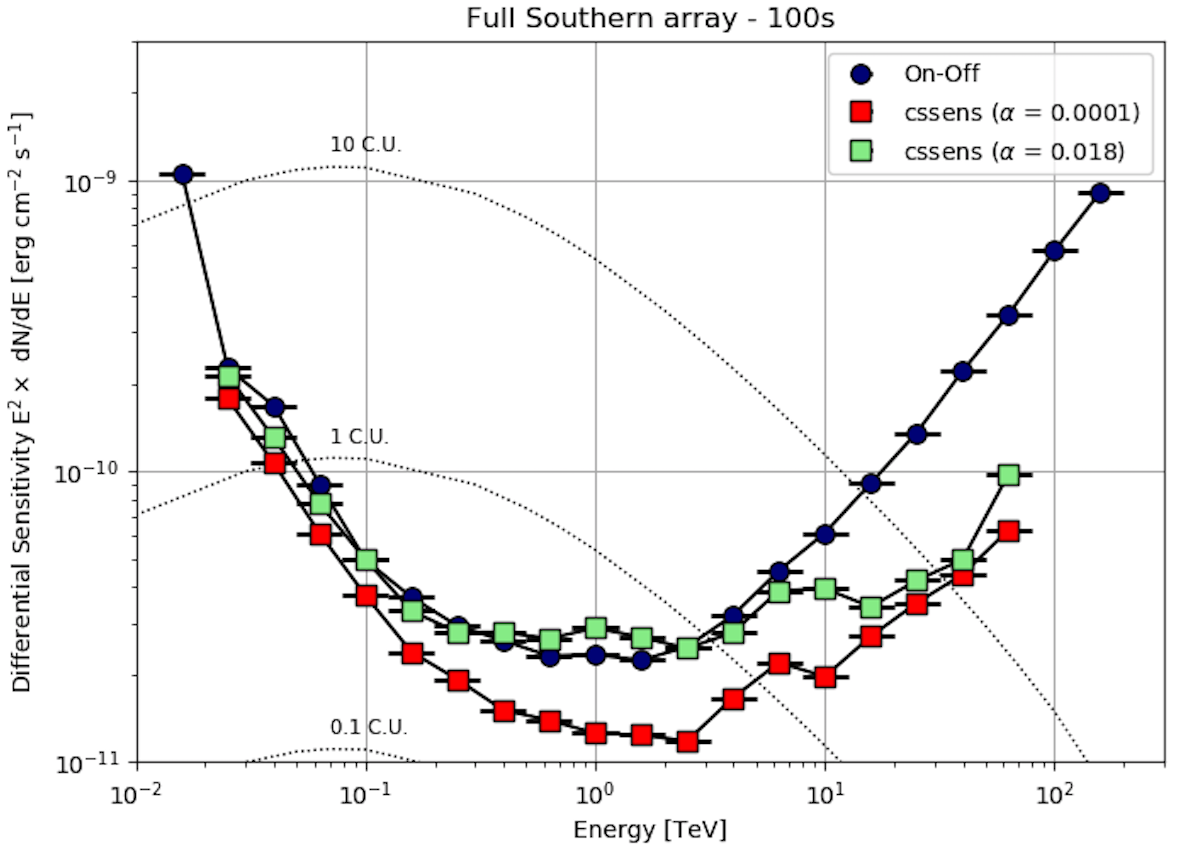}\\
   \includegraphics[width=0.83\textwidth]{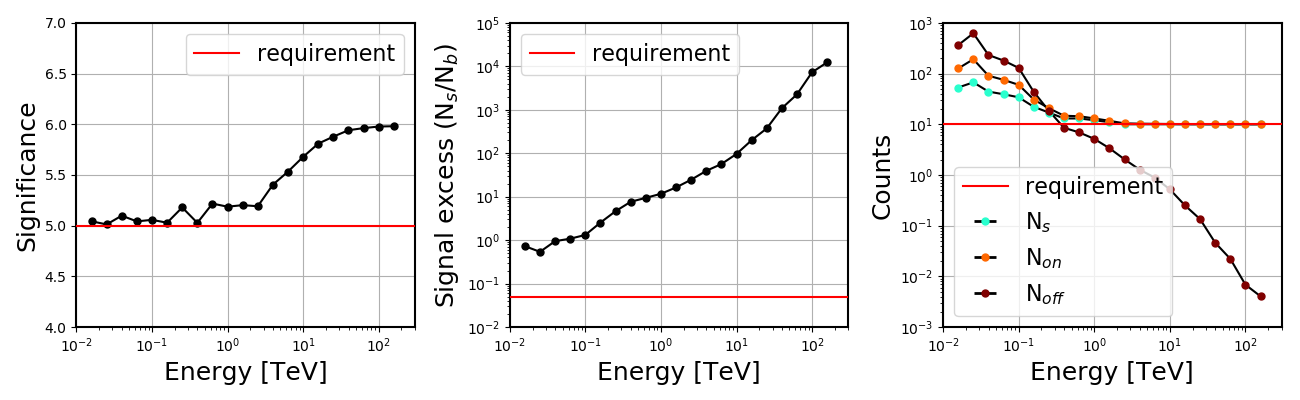}\\
   \caption{\textit{Top panel:} \label{fig:diff_sens_comp}CTA differential sensitivity for a 100 s exposure obtained using the standard \textit{On-Off} procedure and the \textit{ctools/cssens} maximum likelihood method with the baseline and convergence parameters. \textit{Bottom panel:} Spectral behaviour, for the \textit{On-Off} method, of the significance (in equivalent standard deviations, left panel), source flux excess over background (central panel) and \textit{On}, \textit{Off} and source counts (right panel) for a 100 s exposure. The red lines refer to the required value to get a significant detection.}
   \end{figure}
\subsection{Comparing the two methods}
\begin{figure}[t]
   \centering
   \includegraphics[width=0.35\textwidth]{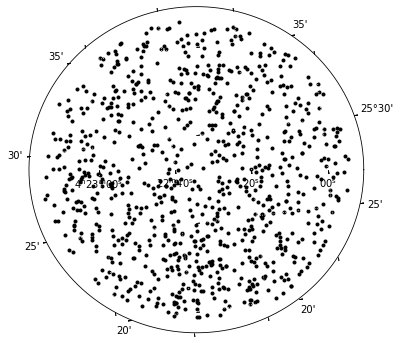}
   \includegraphics[width=0.37\textwidth]{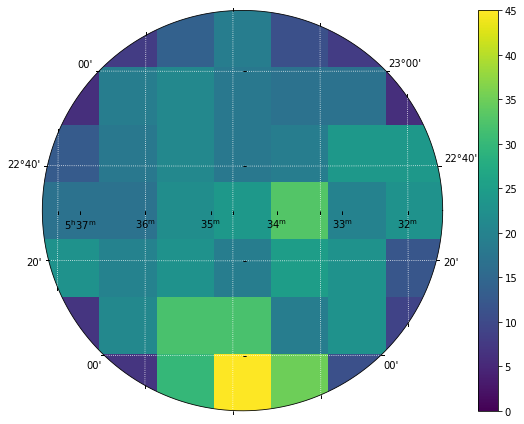}
    \caption{\label{fig:binning}Original, on the left and new, on the right, binned map of the CTA field simulated with \texttt{ctools} for a 100 s exposure in the 19--31 GeV energy bin.} 
   \end{figure}

The energy distribution of the differential sensitivity for the full Southern array \cite{2019APh...111...35A}, for a 100 s exposure and a $20^{\circ}$ zenith angle, is computed using the two methods described above in their standard implementation (Fig. \ref{fig:diff_sens_comp}, top panel).  The CTA Prod3b (June 2017) IRF production was used here, but the relative results would be the same as with the more recent version.
The likelihood modeling technique predicts a better sensitivity than the \textit{On/Off} method along the whole energy range because of the combination of the different assumptions and requirements at its base. 
However, we found that the \texttt{ctlike} test statistic used in \texttt{cssens} can be reduced to the \textit{On-Off} likelihood ratio by an appropriate choice of the sky map binning and simulation region size. The standard simulation region (\ang{5} radius, left panel of Fig. \ref{fig:binning}) implies a strong knowledge of the background, while having small enough pixels to achieve many more background vs signal regions. 
On the other hand, the On-Off approach generally chooses small a priori background and signal regions with angular radius $\theta \sim 0.1\textrm{\textdegree}$ and some geometric observation patterns (Ring Background Model or Wobble) for the selection of the analysis regions. 

If we estimate that the signal region in the simulation is the size of the PSF (Point Spread Function, \SI{68}{\percent} containment radius for selected events dependent on energy), we can label bins within the PSF as signal bins, and all those outside as background bins.
After a trade-off analysis of the \texttt{ctlike} method, we create a $7\times7$ grid (Fig. \ref{fig:binning}, right panel) with larger pixels and reduced radius ($\sqrt{6}\times \rm PSF$) to get a total number of \textit{Off} bins 5 times than the \textit{On} bins (a $3\times3$ grid).
With this configuration, we increase $\alpha$ from $10^{-4}$ to 0.018, close to the one used in the On-Off method.

The resulting sensitivity is compared again to the On-Off result (green squares in Fig \ref{fig:diff_sens_comp}). 
We find notable behaviors at three main energy ranges: a good convergence at \SIlist{25;100}{\GeV} bins, with a slight discrepancy at \SIlist{40;75}{\GeV}; a bump at middle energies around \SI{1}{\TeV}; and no convergence at high energies ($>5$ TeV). 
These differences can be understood if we analyze how the main On-Off parameters (\textit{On} and \textit{Off} counts, the source over the background excess and the significance) impact the sensitivity as shown in Fig. \ref{fig:diff_sens_comp} (bottom panel). At low and medium energies the significance, depending on the $\alpha$ parameter, drives the detection (left panel), hence the convergence with the maximum likelihood technique obtained with the new binning. At higher energies, because of the shortage of gamma-ray photons the ability to collect at least 10 source counts limits the sensitivity in the On-Off method (right panel). Since the \texttt{cssens} script does not require a minimum number of source counts, it reaches significant detections at lower fluxes. At $\sim1$ TeV, the bump is induced by the low statistics in the simulation process because of a low number of values in the fitting process. Calculating the sensitivity multiple times and verifying this distribution, we found that the standard deviation at each bin is fairly wide and could account for the bump. 

Ultimately, we are able to justify the discrepancy between the aperture photometry and the full field maximum likelihood technique. An observer using either method should carefully consider the data-set for proper selection of background and check that the requirements are met (e.g. the minimum number of source counts). 
It should be noted that the number of background counts for a 100 s exposure in the \textit{Off} observation (bottom-right panel of Fig. \ref{fig:diff_sens_comp}) falls well below 10 and tends to zero as we reach the highest energies, violating the requirements of the Li\&Ma equation. In a background-free regime, the $>10$ requirement on the number of source counts would enable a flux statistical uncertainty in each bin $<30\%$. Contrary, if one is interested in "discovery sensitivity" only, the likelihood calculation performed by \texttt{ctools} would still allow a robust detection.


\section{Sensitivity as a function of observation time: CTA, Fermi--LAT, MAGIC, and VERITAS}\label{sec:sens_time}

\begin{figure}[h!]
   \centering
   \includegraphics[width=0.84\textwidth]{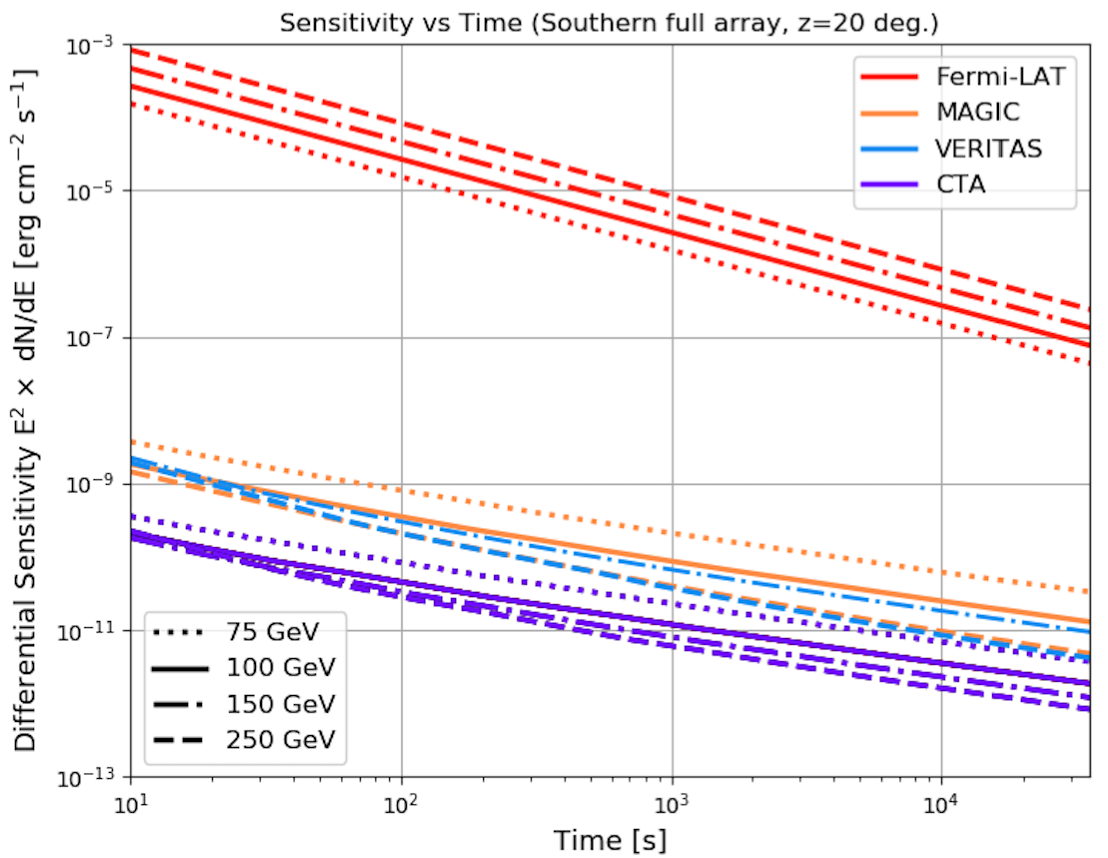}
   \caption{\label{fig:sens_time}CTA differential sensitivity (Southern array, purple lines) as a function of the observation time compared to the \emph{Fermi}-LAT (red lines), MAGIC (orange lines) and VERITAS (light blue lines) sensitivity at four selected energies (75, 100, 150 and 250 GeV).}
   \end{figure}

Following the work of \cite{2013APh....43..348F}, we compute the CTA sensitivity to detect point-like sources at 5 standard deviations significance as a function of the exposure time, for four selected energies. The observation time ranges from 10 s to 8 hours, a full observation night for IACTs. 
Such evolution, shown in Fig. \ref{fig:sens_time}, allows us to explore the CTA ability to detect short-term phenomena and its improvement with respect to Fermi--LAT (Sec. \ref{sec:fermi}), MAGIC (Sec. \ref{sec:magic}) and VERITAS (Sec. \ref{sec:veritas}) sensitivity.
For the CTA sensitivity evaluation, using the full Southern array for a $20^{\circ}$ zenith angle, we make use of the standard \textit{On-Off} method, that predicts a more conservative performance of the array, while keeping in mind the limit of the method at very short exposures. 
   
\subsection{Fermi-LAT}\label{sec:fermi}
The Large Area Telescope (LAT) on board the NASA Fermi gamma-ray observatory \cite{fermi_lat} is a pair-conversion telescope, composed of a Silicon tracker and a CsI calorimeter, covering the energy range from 20 MeV to more than 300 GeV. 
We used \texttt{FermiPy}\footnote{\href{https://fermipy.readthedocs.io}{https://fermipy.readthedocs.io}}, an open source analysis package used by many in the \emph{Fermi}-LAT collaboration, to calculate the flux sensitivity. Due to the well-modeled background of the Fermi Large Area Telescope, the \texttt{FermiPy} flux sensitivity function uses the Asimov method to find the expected normalization, for a given TS threshold and assuming a fixed background model with no uncertainty in the background amplitude.

We used a livetime cube generated from real data and spacecraft file for 6 yrs of observation, with the Pass 8\footnote{\href{http://www.slac.stanford.edu/exp/glast/groups/canda/lat_Performance.htm}{http://www.slac.stanford.edu/exp/glast/groups/canda/lat\_Performance.htm}} IRF (\texttt{"P8R2\_SOURCE\_V6"}). The TS threshold was the standard $\rm TS=25$. The sensitivity spectrum was calculated for 18 energy bins, at 4 bins per decade for \SI{32}{\MeV}-\SI{1}{\TeV}. We assumed a source defined by a power law energy spectrum, with spectral index of -2.5, and set a minimum of 10 excess counts.

Using released values of the \texttt{P8R2\_SOURCE\_V6} sensitivity, we were able to compare our methods to the results for various pointing directions. At the galactic center, the sensitivity values deviated by as much as \SI{40}{\percent}, but this was shown to be a result of a specific galactic diffuse emission spatial binning technique that changed the values by as much as a factor of 2 between spatial bins. Comparing various pointing directions (120,45), (0,90), (0,30) show 5-10\% agreement with published performance results and our computations, while at the energy ranges 25-250 GeV the variations between them are negligible. Fig. \ref{fig:sens_time} shows the sensitivity curves for $(l,b=10,0)$.

\subsection{MAGIC}\label{sec:magic}
MAGIC is a system of two $17\,{\rm m}$-diameter IACTs located at the Observatory Roque de los Muchachos on the Canary island La Palma, Spain. It is designed to perform gamma-ray astronomy in the energy range from $50\,{\rm GeV}$ to above $50\,{\rm TeV}$. 
The differential sensitivity as a function of the observation time presented in this work is estimated using a subset of the same data sample in \cite{aleksic_2016}, consisting of $11\,{\rm h}$ of Crab Nebula observations with zenith angle below $30^{\circ}$, collected between October 2013 and January 2014. 
The orange curves in Figure \ref{fig:sens_time} represent, given an observation time, the differential flux needed at 75, 100 and 250 GeV energy to obtain a significant detection. The latter is defined with the \textit{On-Off} method, applying the same requirements described in Sec. \ref{sec:onoff}. The exposure ratio is fixed to $\alpha=1/3$. The same energy bins defined for CTA are employed. The $N_{\rm on}$ and $N_{\rm off}$ counts depend on the size of the signal and background regions and on the cut on the \textit{hadronness} parameter, expressing for each event the outcome of the gamma-hadron classification. Half of the data sample is used to select the angular and the hadronness cuts that provide, in each energy bin, at each observation time, the minimum fraction of Crab excesses (i.e. the minimum flux in Crab Units) verifying the detection conditions. To obtain an unbiased estimate of the sensitivity, the optimized cuts are then applied to the remaining half of the sample. 

\subsection{VERITAS}\label{sec:veritas}
\vspace{-2mm}
VERITAS is an array of four, ground based, IACTs in southern Arizona, USA \cite{2015ICRC...34..868S}. VERITAS has the performance capability to detect a point source with 1\% of the Crab Nebula flux in 25h, measuring gamma rays with energies from $\sim 100$ GeV up to $\geq 30$ TeV. 

To measure the short-term sensitivity, we used 40 h of Crab Nebula data taken between 2012 and 2014 to measure gamma ray and background rates using a standard "Hillas" analysis, with gamma/hadron separation cuts optimized for a multi-hour exposure. With these rates, we inverted equation 17 from \cite{1983ApJ...272..317L} to find the minimum flux (in units of Crab Nebula flux) required for a 5-$\sigma$ detection (light blue curves in Fig. \ref{sec:sens_time}). While more advanced analysis methods \cite{2017ICRC_veritas_ITM} are starting to be used, their benefits are not reflected here. We compute sensitivities for VERITAS for the same time bins as CTA, for energy bins centered on 150 and 250 GeV, while also maintaining a minimum of ten excess counts per bin and a minimum 5\% background uncertainty.


   

\section{Results and conclusions}
The outcome of CTA observations at very short term exposures is highly influenced by the knowledge and modeling of the background at low energy and by the ability to collect enough photons at the highest energies. 
The \textit{On-Off} aperture photometry and the full-field maximum likelihood techniques give different results in the detection of transient sources, with the latter predicting a sensitivity up to $\sim6$ times better than what is obtained with the standard CTA sensitivity definition.
However, we proved that the more optimistic prediction of the \textit{cssens} maximum likelihood script is given by the finer assumed knowledge of the background and the lack of a requirement on the minimum number of source counts. The two methods can converge if based on the same assumptions, but such choice - and responsibility - is up to the observer.

A problem remains if very few, or no background counts are collected in the \textit{Off} region, a case potentially possible for $<100$ s exposures. Dedicated studies to prove the feasibility of the Li\&Ma equation in case of very few counts and the search for alternative statistical techniques are still needed. 

CTA will ensure, for short-term ($<10^4$ s) gamma-ray emission below 250 GeV, a sensitivity about ten times better than the MAGIC and VERITAS telescopes along the entire time range.
For exposures below 1 hour, CTA will be able to detect sources more than $10^4$ times fainter, in the 75 -- 250 GeV energy range, with respect to Fermi--LAT. For exposures longer than several hours, the lower duty cycle of CTA, limited to night-time observations of the sky region well above the horizon, restricts the efficiency of CTA in monitoring the variable sky. 
\begin{small}
\acknowledgments
{This work was conducted in the context of the CTA Analysis and Simulation Working Group. We gratefully acknowledge financial support from the agencies and organizations listed here\footnote{\href{http://www.cta-observatory.org/consortium_acknowledgments/}{http://www.cta-observatory.org/consortium\_acknowledgments/}}. This paper has gone through internal review by the CTA Consortium.}
\end{small}

\begin{small}

\end{small}

\begin{thebibliography}{99}

\begin{multicols}{2}



\bibitem{2019APh...111...35A} A. {Acharya}, et al.,  \emph{Monte Carlo studies for the optimisation of the Cherenkov Telescope Array layout}, Astrop. Phys., 111, {35-53}, 2019.


\bibitem{aleksic_2016} J. Aleksi\'c, et al. \emph{The major upgrade of the MAGIC telescopes, Part II: A performance study using observations of the Crab Nebula}, \ap, 72, {76-94}, 2016.

\bibitem{fermi_lat} W.~B. Atwood et al. \emph{The Large Area Telescope on the Fermi Gamma-Ray Space Telescope Mission}, \apj, 697, 1071, 2009.

\bibitem{rta2013} A. Bulgarelli, et al., \emph{The Real-Time Analysis of the Cherenkov Telescope Array Observatory}, {Proc. of the 33rd ICRC}, 2013.


\bibitem{2017ICRC_veritas_ITM} J. Christiansen, et al, \emph{Characterization of a Maximum Likelihood Gamma-Ray Reconstruction Algorithm for VERITAS}, {Proc. of the 35th ICRC}, 2017.

\bibitem{2006A&A...457..899A} J. Cortina, et al., \emph{Highlights of the MAGIC telescopes}, Proc. 32nd ICRC, 12, 147, 2011.


\bibitem{2013APh....43..348F} S. Funk, and J.~A. {Hinton}, and {CTA Consortium}, \emph{Comparison of Fermi-LAT and CTA in the region between 10-100 GeV}, {\ap}, 43, 348-355, 2013.




\bibitem{1983ApJ...272..317L} T.P., {Li} and  Y.Q {Ma}, \emph{Analysis methods for results in gamma-ray astronomy}, \apj, 272, 317-324, 1983

\bibitem{gernot_icrc} G. {Maier}, et al.,  \emph{Performance of the Cherenkov Telescope Array}, in these proceedings, 2019.



\bibitem{2015ICRC...34..868S} D. Staszak, et al., \emph{Science Highlights from VERITAS}, {Proc. of the 34th ICRC}, 2015.

\bibitem{CTA1} W. Wild et al., \emph{Cherenkov Telescope Array (CTA): building the world's largest ground-based gamma-ray observatory}, {Proc. of SPIE}, 10700, 2018. 


\end{multicols}



\end{thebibliography}
\end{document}